\documentclass[english,aps,prl,manuscript]{revtex4}
\usepackage[T1]{fontenc}
\usepackage[latin9]{inputenc}
\usepackage{xcolor}
\usepackage{pdfcolmk}
\usepackage{amstext}
\usepackage{amssymb}
\usepackage{graphicx}
\usepackage{esint}
\PassOptionsToPackage{normalem}{ulem}
\usepackage{ulem}

\makeatletter

\providecolor{lyxadded}{rgb}{0,0,1}
\providecolor{lyxdeleted}{rgb}{1,0,0}

\@ifundefined{textcolor}{}
{%
 \definecolor{BLACK}{gray}{0}
 \definecolor{WHITE}{gray}{1}
 \definecolor{RED}{rgb}{1,0,0}
 \definecolor{GREEN}{rgb}{0,1,0}
 \definecolor{BLUE}{rgb}{0,0,1}
 \definecolor{CYAN}{cmyk}{1,0,0,0}
 \definecolor{MAGENTA}{cmyk}{0,1,0,0}
 \definecolor{YELLOW}{cmyk}{0,0,1,0}
}

\makeatother

\usepackage{babel}
\begin{document}

\title{Topological Modes in One Dimensional Solids and Photonic Crystals}

\author{Timothy J. Atherton}

\affiliation{Tufts University, Center for Nanoscopic Physics, Department of Physics
\& Astronomy, 5 Colby Street, Medford, Massachusetts, USA. 02155}

\author{Celia A. M. Butler}

\affiliation{University of Exeter, Electromagnetic and Acoustic Materials Group,
Department of Physics and Astronomy, Stocker Road, Exeter, United
Kingdom. EX4 4QL }

\author{Melita C. Taylor}

\affiliation{University of Exeter, Electromagnetic and Acoustic Materials Group,
Department of Physics and Astronomy, Stocker Road, Exeter, United
Kingdom. EX4 4QL }

\author{Ian R. Hooper}

\affiliation{University of Exeter, Electromagnetic and Acoustic Materials Group,
Department of Physics and Astronomy, Stocker Road, Exeter, United
Kingdom. EX4 4QL }

\author{Alastair P. Hibbins}

\affiliation{University of Exeter, Electromagnetic and Acoustic Materials Group,
Department of Physics and Astronomy, Stocker Road, Exeter, United
Kingdom. EX4 4QL }

\author{J. Roy Sambles}

\affiliation{University of Exeter, Electromagnetic and Acoustic Materials Group,
Department of Physics and Astronomy, Stocker Road, Exeter, United
Kingdom. EX4 4QL }

\author{Harsh Mathur}

\affiliation{Case Western Reserve University, Department of Physics, 10900 Euclid
Avenue, Cleveland, Ohio, USA 44106}
\begin{abstract}
It is shown theoretically that a one-dimensional crystal with time
reversal symmetry is characterized by a $Z_{2}$ topological invariant
that predicts the existence or otherwise of edge states. This is confirmed
experimentally through the construction and simulation of a photonic
crystal analogue in the microwave regime. 
\end{abstract}
\maketitle
Topological phases have been shown to arise in a number of condensed
matter systems: in the quantum Hall effect\cite{Thouless82} where
electrons are confined to two dimensions and subject to a perpendicular
applied magnetic field and in so-called topological insulators\cite{Hasan:2010p4090,RevModPhys.83.1057}
which are materials the possess conducting metallic surfaces despite
being insulators in the bulk. Experimental studies of these states,
in materials such as graphene\cite{RevModPhys.81.109} and Bi$_{2}$Se$_{3}$
have recently been an area of considerable focus both for fundamental
reasons, because the topological states in these materials lead to
exotic quasi-particles, and also for applications such as quantum
computing. 

The states arise in these systems as follows: consider a map from
the Brillouin zone to a space of nondegenerate Bloch Hamiltonians
$H(k)$. If $\left|k\right\rangle $ is an eigenstate of $H(k)$,
then a vector potential $A(k)=-i\left\langle k\left|\partial_{k}\right|k\right\rangle $
may be defined following Berry\cite{Berry08031984}. The topological
index of this map called the \emph{Chern number} is $Q=\int\nabla\times A(k)\ \text{d}^{2}k$.
Thouless et al. discovered that non-trivial Chern numbers can arise
in the Brillouin zone with time reversal symmetry broken by the application
of a strong magnetic field as in the quantum Hall effect\cite{Thouless82}.
More recently Balents and Moore\cite{Moore:2007p2922} applied this
paradigm to systems with strong spin-orbit interaction but with time
reversal symmetry intact and thereby clarified an earlier proposition
by Kane and Mele\cite{Kane:2005p2973} that a $\mathbb{Z}_{2}$ invariant
of the band structure divides insulators into two classes: an even
class corresponding to conventional insulators and an odd topological
insulating phase that possesses conductive surface states. Both spin-orbit
coupling and breaking of inversion symmetry are prerequisites for
such materials.

In this paper, we apply this paradigm to a system with different symmetry,
a \emph{one}-dimensional crystal with time reversal and charge conjugation
symmetry. In the language of random matrix theory\cite{PhysRevB.55.1142},
our system corresponds to class BDI in contrast to the unitary class
A in the work of Thouless et al.\cite{Thouless82} and the symplectic
class AII in Balents and Moore\cite{Moore:2007p2922}. By an analogous
argument, it is shown that such a structure may possess edge states
characterized by a $\mathbb{Z}_{2}$ topological invariant. The system
is then realized experimentally through a photonic analogue. We stress
that while edge states such as Tamm states are well known in 1D crystals\cite{Fang:2013p4091},
it is the topologically-protected nature of those presented here that
primarily concerns us. 

\emph{Model\textemdash }We begin with the topological argument. Consider
a tight binding model on a one dimensional-lattice with nearest neighbor
hopping and a constant on-site energy. Such a model is described by
the following Schr\"odinger equation,
\begin{equation}
-\tau_{n}\psi_{n+1}-\tau_{n-1}\psi_{n-1}=E\psi_{n},\label{eq:schrodingereqn}
\end{equation}
where as usual the $\psi_{n}$ represent the amplitude of the wavefunction
at the $n$-th lattice site and $\tau_{n}$ is the real hopping coefficient
between the $n$-th site to the $n+1$th site. As is well-known, for
arbitrary $\tau_{n}$, the model (\ref{eq:schrodingereqn}) possesses
a time-reversal symmetry with associated operator, 
\begin{equation}
\mathcal{T}\psi_{n}=\psi_{n}^{*},
\end{equation}
as well as a particle-hole $\mathcal{C}$ symmetry represented by
the anti-unitary charge conjugation operator,
\begin{equation}
\mathcal{C}\psi_{n}=(-1)^{n}\psi_{n}^{*},\label{eq:chargeconjugation}
\end{equation}
and where,
\begin{equation}
\mathcal{C}^{2}=1,\ \mathcal{T}^{2}=1.
\end{equation}
The particle-hole symmetry restricts the Hamiltonian's spectrum because
if $\left|\psi\right\rangle $ is an eigenfunction of (\ref{eq:schrodingereqn})
with energy $E$, then $\mathcal{C}\left|\psi\right\rangle $ is also
an eigenfunction with energy $-E$ . 

For simplicity, a bi-partite lattice is now considered where the bond
strengths are alternately $\tau_{1}$ and $\tau_{2}$. The Schrödinger
equation for this situation has the form 
\begin{eqnarray}
-\tau_{1}\phi_{n}^{B}-\tau_{2}\phi_{n-1}^{B} & = & E\phi_{n}^{A},\nonumber \\
-\tau_{1}\phi_{n}^{A}-\tau_{2}\phi_{n+1}^{A} & = & E\phi_{n}^{B}.\label{eq:bipartiteschrodinger}
\end{eqnarray}
This may be diagonalized by introducing the Bloch ansatz, $\phi_{n}^{A}=\alpha e^{ikn},\;\phi_{n}^{B}=\beta e^{ikn}$,
yielding, 
\begin{equation}
\left(\begin{array}{cc}
0 & z^{\ast}(w)\\
z(w) & 0
\end{array}\right)\left(\begin{array}{c}
\alpha\\
\beta
\end{array}\right)=E\left(\begin{array}{c}
\alpha\\
\beta
\end{array}\right),\label{eq:blochham}
\end{equation}
where the $2\times2$ matrix on the left of eq (\ref{eq:blochham})
is the Bloch Hamiltonian, $w=e^{ik}$ and $z(w)=-(\tau_{1}+\tau_{2}w)$.
Eq (\ref{eq:blochham}) thus defines a map from the Brillouin zone
$-\pi\leq k<\pi$ to the space of Bloch Hamiltonians. Equivalently,
this may be viewed as a map $z(w)$ from a unit circle in the $w$
plane (the Brillouin zone) to the complex plane with the origin excluded.
The origin is excluded provided, as we assume, the bands are non-degenerate.
The $\pi_{1}$ homotopy of the punctured plane is well known to be
non-trivial. The Chern number $Q$ in this case corresponds to the
number of times the loop $z(e^{ik})$ winds about the origin. It is
easy to see that for the case $\tau_{1}<\tau_{2}$, $Q=1$ {[}fig.
1(a)(i){]}; for $\tau_{1}>\tau_{2}$, $Q=0$ {[}fig. 1(a)(ii){]}.
This can also be determined more formally by constructing the eigenspinors
of the Bloch Hamiltonian, computing the corresponding Berry connection,
and evaluating the Wilson loop $\int dkA(k)$. We will explain below
why higher values of $Q$ cannot be obtained even if we perturb the
model (e.g. by incorporating longer range hopping that respects ${\cal C}$
and ${\cal T}$ symmetry).

It is straightforward to verify that the Chern number is associated
with a zero energy bound state by noting that if the lattice is curtailed,
so that $n=0$ represents the leftmost edge, the wave function will
obey the boundary condition 
\begin{eqnarray}
-\tau_{1}\phi_{0}^{B} & = & E\phi_{0}^{A},\nonumber \\
-\tau_{1}\phi_{0}^{A}-\tau_{2}\phi_{1}^{A} & = & E\phi_{0}^{B}.\label{eq:boundary}
\end{eqnarray}
Setting $E=0$ it follows that the solution to eqs (\ref{eq:bipartiteschrodinger})
and (\ref{eq:boundary}) is 
\begin{eqnarray}
\phi_{n}^{B} & = & 0,\nonumber \\
\phi_{n}^{A} & = & (-\tau_{1}/\tau_{2})^{n}\phi_{0}^{A},\label{eq:edgemode}
\end{eqnarray}
which is finite as $n\rightarrow+\infty$ only if $\tau_{1}<\tau_{2}$,
and in such a case is manifestly localized in character.

Now let us reformulate the argument with more generality. We continue
to assume that the crystal has a bi-partite lattice and that the Hamiltonian
commutes with ${\cal T}$ and anti-commutes with ${\cal C}$. ${\cal C}$
and ${\cal T}$ transform a plane wave of wave vector $k$ into one
of wave-vector $-k$. At the same time the amplitudes $(\alpha,\beta)$
are transformed to $(\alpha^{\ast},-\beta^{\ast})$ and $(\alpha^{\ast},\beta^{\ast})$
respectively. The two symmetries relate the Bloch Hamiltonians at
wave-vectors $k$ and $-k$ via $H(-k)={\cal T}H(k){\cal T}=-{\cal C}H(k){\cal C}$.
Together these relations and the requirement of hermiticity constrain
the $2\times2$ Bloch Hamiltonian $H(k)$ to be off-diagonal. At the
special points $k=0$ and $k=\pi$ which are invariant under $k\rightarrow-k$
the Bloch Hamiltonian is required to have the form 
\begin{equation}
H=\left(\begin{array}{cc}
0 & -ia\\
ia & 0
\end{array}\right).\label{eq:blocha}
\end{equation}
Here $a$ is real and $a\neq0$ since we are assuming that the bands
have no accidental degeneracies. Topologically, this space is a punctured
line. Rather than considering the mapping of the entire Brillouin
zone to the space of Bloch Hamiltonians as before we may instead consider
the mapping from the two special points $k=0$ and $k=\pi$ to the
punctured line. Such maps fall into two classes: a trivial one {[}fig.
1(b)(i){]} where the special points are mapped to the same side of
the origin, $a=0$, and a non-trivial case {[}fig. 1(b)(ii){]} such
that the special points are mapped to opposite sides%
\footnote{If we assume that for generic values of $k$ the Bloch Hamiltonian
will not accidentally have the form eq (\ref{eq:blocha}) then the
map from the Brillouin zone to the complex plane can cross the $x$
axis only at $k=0$ and $k=\pi$; this is the reason that the winding
number can only equal $0$ or $1$.%
}. The edge mode occurs in the non-trivial case.

Our argument mirrors that of Balents and Moore\cite{Moore:2007p2922}
who considered maps from an effective Brillouin zone in two dimensions
which had the topology of a cylinder to the space of Bloch Hamiltonians.
The circular boundaries of the effective Brillouin zone in their argument
are analogous to our special points $k=0$ and $\pi$. Apart from
the difference in dimensionality another key difference between their
work and ours is in the symmetry. As noted above they considered bands
with odd time reversal symmetry (the symplectic class AII) whereas
we consider even time reversal symmetry accompanied by charge conjugation
symmetry (class BDI). As a consequence breaking of spatial inversion
symmetry is essential for non-trivial topology in their analysis but
not for the class of topological insulator considered here. Indeed
the model eq (1) respects parity.

We now address the issue of how the edge state might be observed.
The hopping model on a bipartite lattice is actualized in a 1D solid
where an electron is subject to a periodic array of potential barriers
of alternating height. The reflection and transmission of incident
free particle wavefunctions of wavevector $k$ through such a structure
is readily determined by matching local solutions to the Schr\"odinger
equation at interfaces using matrix methods\cite{Sprung:2007p3124}.
To do so, define, 
\begin{equation}
T(\theta)=\left(\begin{array}{cc}
e^{i\chi}\cosh\theta & \sinh\theta\\
\sinh\theta & e^{-i\chi}\cosh\theta
\end{array}\right),\ U=\left(\begin{array}{cc}
e^{ikb} & 0\\
0 & e^{-ikb}
\end{array}\right).\label{eq:transfermatrices}
\end{equation}
Here $T(\theta)$ is the transfer matrix for a single symmetric barrier
located at the origin and $U$ a translation operator; $\theta$ is
the opacity of the barrier, $b$ is the lattice spacing. In the limit
of large barriers, $\chi=ka+\xi$ where $a$ is the barrier width
and $\xi$ is an overall phase shift. 

The band structure of the bipartite lattice, with alternating barriers
of opacity $\theta_{1}$ and $\theta_{2}$ respectively, is determined
by finding the eigenvalues of the transfer matrix corresponding to
the unit cell, i.e. $UT(\theta_{1})UT(\theta_{2})$. For $\theta_{2}\neq\theta_{1}$,
the usual band, i.e. values of $k$ such that the eigenvalues are
complex, is found to split into two sub-bands symmetrically placed
around the point $k_{0}=(\pi/2-\xi)/(a+b)$ and with edges corresponding
to the roots of $\cos(2k(a+b)+2\xi)\cosh\theta_{1}\cosh\theta_{2}+\sinh\theta_{1}\sinh\theta_{2}=\pm1$.
Between these subbands, including the point $k_{0}$, the reflection
coefficient $|r|^{2}\to1$; the phase shift $\delta$ of a reflected
wave, however, experiences a jump of $\pi$ around $k_{0}$ if $\theta_{1}<\theta_{2}$
and $0$ otherwise. The jump leads to a Lorentzian feature in the
time delay $\frac{\partial\delta}{\partial k}$ of a reflected wave;
the width was determined by routine calculations to be,
\begin{equation}
L=2e^{\theta_{2}}\cosh\theta_{1}\frac{\sinh(\theta_{1}+\theta_{2})-\cosh\theta_{1}\cosh\theta_{2}}{\sinh^{2}(\theta_{2}-\theta_{2})}.
\end{equation}
Hence, the topological mode may be observed experimentally by a technique
such as Time Delay Reflection Spectroscopy. As will be seen shortly,
for a finite structure evidence of the topological mode is also available
in some circumstances from the reflection profile. 

To verify the above arguments, we numerically evaluated the reflection
coefficient and time delay for a 3.5 period structure from the transfer
matrix as a function of $k$ and $\theta_{2}$ with fixed $\theta_{1}=1$,
$a=0.1$, $b=1$. Results are plotted in fig. \ref{fig:varythickness}.
When $\theta_{1}=\theta_{2}$ the reflection spectrum {[}fig. \ref{fig:varythickness}(a){]}
consists of 6 minima, which correspond to a band in a structure with
an infinite number of unit cells. As $\theta_{2}$ is increased, the
central two minima move closer together and merge around $\theta_{2}\approx1.3$;
as $\theta_{2}\to\infty$, this mid-band mode vanishes in reflection,
contributing only a phase shift to the reflected light as discussed
earlier. Conversely, if $\theta_{2}<\theta_{1}$, the central minima
become separated and, as predicted by the topological argument above,
no mid-band mode exists. Results for 6.5 units are shown in fig. \ref{fig:varythickness}(c)
and (d); the additional layers generate corresponding modes, but the
topological mode remains for $\theta_{2}>\theta_{1}$. If the number
of periods is increased further, more modes appear to eventually yield
the two sub-bands of the infinite bipartite lattice and the mid-gap
mode persists confirming its topological nature.

\emph{Photonic analogue\textemdash{}} In order to conveniently verify
the veracity of the topological index as a predictor of edge states,
we exploit the known isomorphism between the Schr\"odinger equation
and Maxwell's equations in one dimension\cite{Hooper2006} to construct
a photonic analogue structure in the microwave regime.

The experimental setup is illustrated in fig. \ref{fig:Expt} and
details of the data collection described more fully in \cite{butler}.
The experimental structure itself consists of alternating layers of
dielectric (air) and metamaterials. The metamaterial layers are separated
by metallic spacers around the edge of the sample yielding a nominal
spacing of $t_{air}=7.60\pm0.01\text{\ mm}$. The metamaterial layers
are made from a solid aluminium sheet perforated with a periodic square
array of holes of pitch $d=7.68\pm0.01\ \text{mm}$ and hole size
of $l=6.15\pm0.01\ \textrm{mm}$; the thickness of the metamaterial
in successive layers is alternated, and denoted $t_{A}$ and $t_{B}$
respectively, to form the required ABAB bipartite stack with 3.5 unit
cells as modelled earlier. When illuminated with microwave radiation
between $12$ and $20$ GHz, the metamaterial supports only evanescently
decaying modes in the subwavelength holes; hence the metamaterial
layer behaves as an \emph{effective} metal layer \cite{McCalmont}
with a skin depth of $\sim1\ \text{mm}$. This arrangement is used
rather than homogenous metal plates because metals behave as near
perfect electrical conductors at microwave frequencies; thin metallic
films could be used but by using a metamaterial the resulting structure
is less sensitive to variations in thickness. The effective properties
of the metamaterial layer can also be controlled much more precisely
by adjusting the design parameters i.e. thickness, pitch and fill
fraction.

In the experiment, the sample is illuminated by a collimated $s$-polarized
(transverse electric) microwave beam, produced by a microwave horn
placed at the focus of a spherical mirror, incident on the sample
at $10^{\circ}.$ Reflected intensity is measured as a function of
frequency using a detector horn and secondary mirror. The sample studied
had $t_{A}<t_{B}$, specifically $t_{A}=0.66\pm0.01\ \text{mm}$ and
$t_{B}=2.33\pm0.01\ \text{mm}$ as measured from the constructed sample.
Since the opacity parameters $\theta$ of eq. (\ref{eq:transfermatrices})
monotonically increase with the thicknesses $t$, the topological
argument above predicts that the edge states should be observed in
this case. The reflected intensity is plotted as a function of frequency
in fig. \ref{fig:expt}(a), confirming the presence of the midgap
topological mode. Due to the finite number of periods in the structure,
the band is manifested as a series of reflection minima corresponding
to resonant modes in the structure; if the number of unit cells were
increased, the continuous sub-bands would be recovered\cite{Griffiths:2001p2889}.
We also simulated the response of a reversed structure, i.e. where
$t_{A}=2.33\ \text{mm}$ and $t_{B}=0.66\ \text{mm}$; the reflection
profile is plotted in gray in fig. \ref{fig:expt}(a). No mid-gap
topological mode is observed in this structure in agreement with the
above prediction. 

To characterize the nature of these modes, we numerically modelled
the reflection response of the structures using commercial finite-element
software (Comsol). As with other experiments\cite{butler09,butler11},
modelling can be used to visualize the electric field distribution
in each mode. The simulated response is plotted in fig. \ref{fig:expt}(a).
We emphasize that no fitting was performed and the agreement between
model and data displayed in fig. \ref{fig:expt}(a) is typical for
such studies, accurately locating the position of the minima while
incorrectly estimating their depth. This discrepency is due to a number
of physical effects including: finite area of the sample in the experiment;
spherical aberration of the microwave source and detector; imperfections
in the experimental sample such as bowing of the metamaterial layers
which changes the dielectric spacing; and a radius of curvature associated
with the hole edges that reduces the effective skin depth.

Plots of the time averaged electric field magnitude as a function
of the distance through the sample are displayed in fig. \ref{fig:expt}(b)
for each reflection minimum in the fitted profile fig. \ref{fig:expt}(a).
Modes A,B,D and E are clearly distinct in character from mode C. Mode
C has the predicted properties of the topological edge state: it is
confined to the edge of the sample and also occurs in the middle of
the band. Plots of the average electric field are also shown in fig.
\ref{fig:expt}(c) for the six minima in the reversed structure. Unlike
the previous case, none of the modes i\textemdash vi are localized
to the edges, in agreement with the prediction of the above topological
argument. 

\emph{Conclusion\textemdash }A topologically protected edge state
has been experimentally observed in a 1D photonic crystal with time-reversal
symmetry. The existence or otherwise of the state as a function of
the design parameters of the crystal is predicted by a $\mathbb{Z}_{2}$
topological invariant that classifies mappings from the band structure
to the space of Bloch Hamiltonians; the classification was constructed
using methods previously applied to two and three dimensional topological
insulators. However the crystals we consider are of a different class
from conventional topological insulators in terms of their symmetries.
The latter must have time reversal symmetry, strong spin-orbit interaction
and a lack of inversion symmetry; in our work we use an anti-unitary
charge conjugation symmetry $\mathcal{C}$ as well as the time-reversal
operator $\mathcal{T}$ to develop the classification. 

\emph{Acknowledgement\textemdash The authors wish to acknowledge financial
support from the EPSRC through the QUEST programme grant (EP/I034548/1)
and also QinetiQ and the EPSRC for funding CAMB through the Industrial
CASE scheme (no. 08000346). HM is supported by a grant from the U.S.
Department of Energy to the Particle Astrophysics Theory group at
CWRU. }

\newpage{}

\begin{figure}
\includegraphics{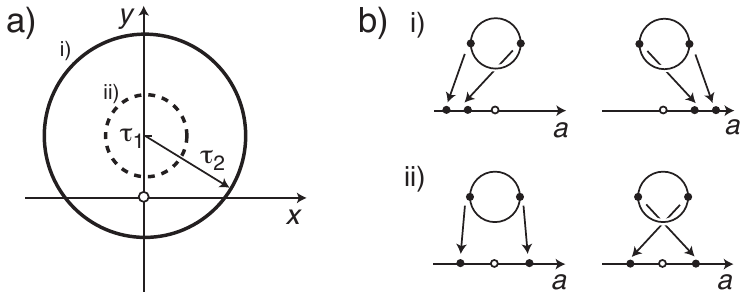}\protect\caption{\label{fig:Maps}(a) Maps from the Brillouin zone to the space of
$2\times2$ Hamiltonians (\ref{eq:blochham}) fall into 2 classes
depending on the value of parameters $\tau_{1}$ and $\tau_{2}$:
i) those that encircle the origin and ii) those that do not. (b) Maps
from the Brillouin zone to the space of $C$-symmetric Hamiltonians
are i) trivial if both points are mapped to the same side of the excluded
point $a=0$ and ii) nontrivial otherwise. }
\end{figure}

\begin{figure}
\begin{centering}
\includegraphics{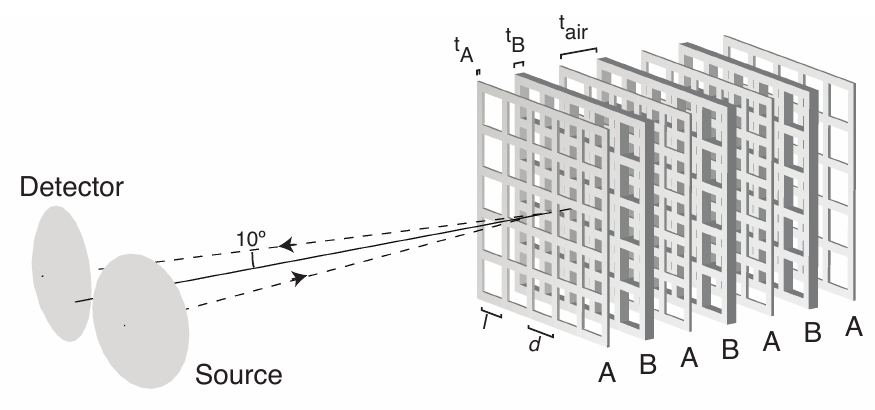}
\par\end{centering}

\protect\caption{\label{fig:Expt}Sketch of experimental setup to obtain reflected
intensity profiles of microwave radiation from the metamaterial structure.
The structure consists of metamaterial layers of alternating thicknesses
$t_{A}=0.66\ \text{mm}$ and $t_{B}=2.33\ \text{mm}$. The metamaterial
layers are aluminium plates stamped with a square array of pitch $d=7.68\ \text{mm}$
and hole size $l=6.15\ \text{mm}$; the metamaterials are spaced by
air with $t_{air}=7.6\text{mm}$.}
\end{figure}

\begin{figure}
\includegraphics{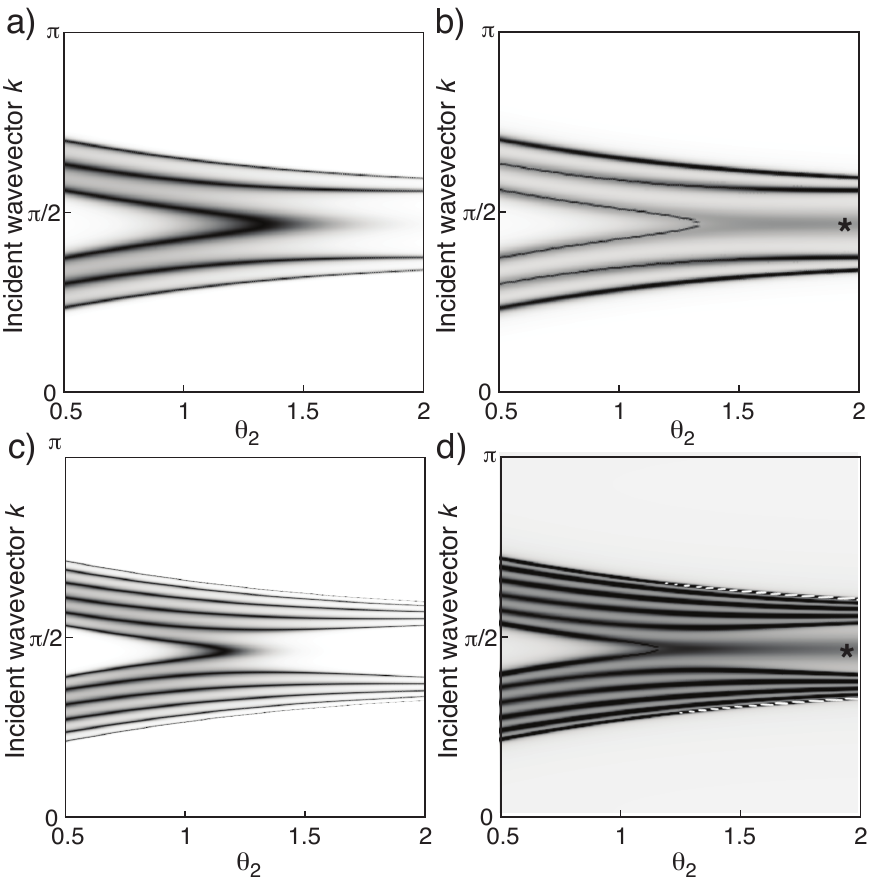}

\protect\caption{\label{fig:varythickness}(a) Reflected intensity as a function of
wavevector for opacity parameters $\theta_{1}=1$ and varying $\theta_{2}$.
(b) Time delay as a function of wavevector and $\theta_{2}$; the
lorentzian feature indicative of the topological mode for $\theta_{2}>\theta_{1}$
is indicated with a {*}. Corresponding plots are shown for 6.5 units
in (c) and (d).}
\end{figure}

\begin{figure}
\begin{centering}
\includegraphics{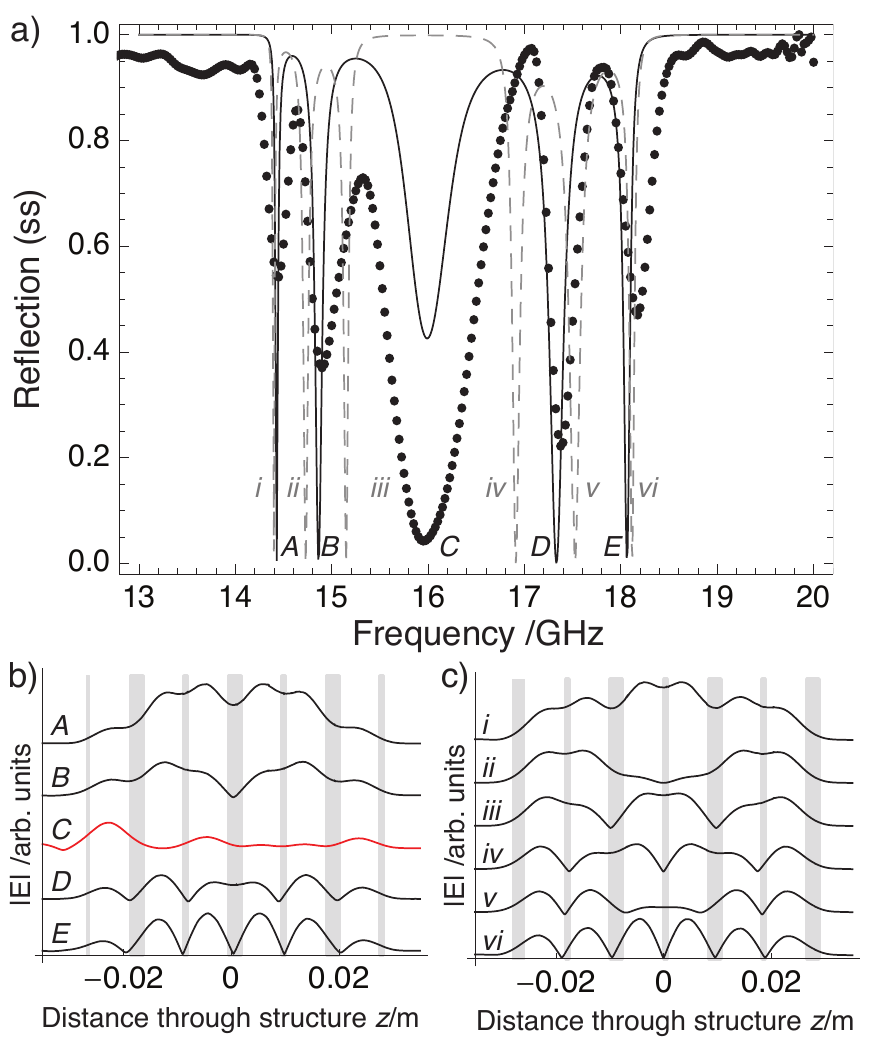}
\par\end{centering}

\centering{}\protect\caption{\label{fig:expt}(a) Measured (disks) and modelled (black solid line)
reflection profile for a 3.5 period AB bipartite stack with $t_{A}<t_{B}$;
reflection minima are labelled (A-E). Also plotted (grey dashed line)
is the reflection profile for a structure with $t_{A}>t_{B}$ with
reflection minima labelled (i-vi). Incident radiation is from the
left. (b) Normalized electric field intensity profile for each observed
mode A-E, calculated from finite-element modelling of the experimental
setup. Position of metamaterial layers is indicated by gray shading.
The topological mode C is highlighted in red. (c) Corresponding plots
for modes of the reversed structure i-vi. }
\end{figure}


\begin{thebibliography}{16}%
\makeatletter
\providecommand \@ifxundefined [1]{%
 \@ifx{#1\undefined}
}%
\providecommand \@ifnum [1]{%
 \ifnum #1\expandafter \@firstoftwo
 \else \expandafter \@secondoftwo
 \fi
}%
\providecommand \@ifx [1]{%
 \ifx #1\expandafter \@firstoftwo
 \else \expandafter \@secondoftwo
 \fi
}%
\providecommand \natexlab [1]{#1}%
\providecommand \enquote  [1]{``#1''}%
\providecommand \bibnamefont  [1]{#1}%
\providecommand \bibfnamefont [1]{#1}%
\providecommand \citenamefont [1]{#1}%
\providecommand \href@noop [0]{\@secondoftwo}%
\providecommand \href [0]{\begingroup \@sanitize@url \@href}%
\providecommand \@href[1]{\@@startlink{#1}\@@href}%
\providecommand \@@href[1]{\endgroup#1\@@endlink}%
\providecommand \@sanitize@url [0]{\catcode `\\12\catcode `\$12\catcode
  `\&12\catcode `\#12\catcode `\^12\catcode `\_12\catcode `\%12\relax}%
\providecommand \@@startlink[1]{}%
\providecommand \@@endlink[0]{}%
\providecommand \url  [0]{\begingroup\@sanitize@url \@url }%
\providecommand \@url [1]{\endgroup\@href {#1}{\urlprefix }}%
\providecommand \urlprefix  [0]{URL }%
\providecommand \Eprint [0]{\href }%
\providecommand \doibase [0]{http://dx.doi.org/}%
\providecommand \selectlanguage [0]{\@gobble}%
\providecommand \bibinfo  [0]{\@secondoftwo}%
\providecommand \bibfield  [0]{\@secondoftwo}%
\providecommand \translation [1]{[#1]}%
\providecommand \BibitemOpen [0]{}%
\providecommand \bibitemStop [0]{}%
\providecommand \bibitemNoStop [0]{.\EOS\space}%
\providecommand \EOS [0]{\spacefactor3000\relax}%
\providecommand \BibitemShut  [1]{\csname bibitem#1\endcsname}%
\let\auto@bib@innerbib\@empty
\bibitem [{\citenamefont {Thouless}\ \emph {et~al.}(1982)\citenamefont
  {Thouless}, \citenamefont {Kohmoto}, \citenamefont {Nightingale},\ and\
  \citenamefont {den Nijs}}]{Thouless82}%
  \BibitemOpen
  \bibfield  {author} {\bibinfo {author} {\bibfnamefont {D.~J.}\ \bibnamefont
  {Thouless}}, \bibinfo {author} {\bibfnamefont {M.}~\bibnamefont {Kohmoto}},
  \bibinfo {author} {\bibfnamefont {M.~P.}\ \bibnamefont {Nightingale}}, \ and\
  \bibinfo {author} {\bibfnamefont {M.}~\bibnamefont {den Nijs}},\ }\href@noop
  {} {\bibfield  {journal} {\bibinfo  {journal} {Phys. Rev. Lett.}\ }\textbf
  {\bibinfo {volume} {49}},\ \bibinfo {pages} {405} (\bibinfo {year}
  {1982})}\BibitemShut {NoStop}%
\bibitem [{\citenamefont {Hasan}\ and\ \citenamefont
  {Kane}(2010)}]{Hasan:2010p4090}%
  \BibitemOpen
  \bibfield  {author} {\bibinfo {author} {\bibfnamefont {M.~Z.}\ \bibnamefont
  {Hasan}}\ and\ \bibinfo {author} {\bibfnamefont {C.~L.}\ \bibnamefont
  {Kane}},\ }\href@noop {} {\bibfield  {journal} {\bibinfo  {journal} {Rev.
  Mod. Phys.}\ }\textbf {\bibinfo {volume} {82}},\ \bibinfo {pages} {3045}
  (\bibinfo {year} {2010})}\BibitemShut {NoStop}%
\bibitem [{\citenamefont {Qi}\ and\ \citenamefont
  {Zhang}(2011)}]{RevModPhys.83.1057}%
  \BibitemOpen
  \bibfield  {author} {\bibinfo {author} {\bibfnamefont {X.-L.}\ \bibnamefont
  {Qi}}\ and\ \bibinfo {author} {\bibfnamefont {S.-C.}\ \bibnamefont {Zhang}},\
  }\href@noop {} {\bibfield  {journal} {\bibinfo  {journal} {Rev. Mod. Phys.}\
  }\textbf {\bibinfo {volume} {83}},\ \bibinfo {pages} {1057} (\bibinfo {year}
  {2011})}\BibitemShut {NoStop}%
\bibitem [{\citenamefont {Castro~Neto}\ \emph {et~al.}(2009)\citenamefont
  {Castro~Neto}, \citenamefont {Guinea}, \citenamefont {Peres}, \citenamefont
  {Novoselov},\ and\ \citenamefont {Geim}}]{RevModPhys.81.109}%
  \BibitemOpen
  \bibfield  {author} {\bibinfo {author} {\bibfnamefont {A.~H.}\ \bibnamefont
  {Castro~Neto}}, \bibinfo {author} {\bibfnamefont {F.}~\bibnamefont {Guinea}},
  \bibinfo {author} {\bibfnamefont {N.~M.~R.}\ \bibnamefont {Peres}}, \bibinfo
  {author} {\bibfnamefont {K.~S.}\ \bibnamefont {Novoselov}}, \ and\ \bibinfo
  {author} {\bibfnamefont {A.~K.}\ \bibnamefont {Geim}},\ }\href@noop {}
  {\bibfield  {journal} {\bibinfo  {journal} {Rev. Mod. Phys.}\ }\textbf
  {\bibinfo {volume} {81}},\ \bibinfo {pages} {109} (\bibinfo {year}
  {2009})}\BibitemShut {NoStop}%
\bibitem [{\citenamefont {Berry}(1984)}]{Berry08031984}%
  \BibitemOpen
  \bibfield  {author} {\bibinfo {author} {\bibfnamefont {M.~V.}\ \bibnamefont
  {Berry}},\ }\href@noop {} {\bibfield  {journal} {\bibinfo  {journal} {Proc.
  Roy. Soc. A}\ }\textbf {\bibinfo {volume} {392}},\ \bibinfo {pages} {45}
  (\bibinfo {year} {1984})}\BibitemShut {NoStop}%
\bibitem [{\citenamefont {Moore}\ and\ \citenamefont
  {Balents}(2007)}]{Moore:2007p2922}%
  \BibitemOpen
  \bibfield  {author} {\bibinfo {author} {\bibfnamefont {J.~E.}\ \bibnamefont
  {Moore}}\ and\ \bibinfo {author} {\bibfnamefont {L.}~\bibnamefont
  {Balents}},\ }\href@noop {} {\bibfield  {journal} {\bibinfo  {journal} {Phys.
  Rev. B}\ }\textbf {\bibinfo {volume} {75}},\ \bibinfo {pages} {121306}
  (\bibinfo {year} {2007})}\BibitemShut {NoStop}%
\bibitem [{\citenamefont {Kane}\ and\ \citenamefont
  {Mele}(2005)}]{Kane:2005p2973}%
  \BibitemOpen
  \bibfield  {author} {\bibinfo {author} {\bibfnamefont {C.~L.}\ \bibnamefont
  {Kane}}\ and\ \bibinfo {author} {\bibfnamefont {E.~J.}\ \bibnamefont
  {Mele}},\ }\href@noop {} {\bibfield  {journal} {\bibinfo  {journal} {Phys.
  Rev. Lett.}\ }\textbf {\bibinfo {volume} {95}},\ \bibinfo {pages} {146802}
  (\bibinfo {year} {2005})}\BibitemShut {NoStop}%
\bibitem [{\citenamefont {Altland}\ and\ \citenamefont
  {Zirnbauer}(1997)}]{PhysRevB.55.1142}%
  \BibitemOpen
  \bibfield  {author} {\bibinfo {author} {\bibfnamefont {A.}~\bibnamefont
  {Altland}}\ and\ \bibinfo {author} {\bibfnamefont {M.~R.}\ \bibnamefont
  {Zirnbauer}},\ }\href@noop {} {\bibfield  {journal} {\bibinfo  {journal}
  {Phys. Rev. B}\ }\textbf {\bibinfo {volume} {55}},\ \bibinfo {pages} {1142}
  (\bibinfo {year} {1997})}\BibitemShut {NoStop}%
\bibitem [{\citenamefont {tuan Fang}\ \emph {et~al.}(2013)\citenamefont {tuan
  Fang}, \citenamefont {kun Chen}, \citenamefont {Zhu},\ and\ \citenamefont
  {Zhou}}]{Fang:2013p4091}%
  \BibitemOpen
  \bibfield  {author} {\bibinfo {author} {\bibfnamefont {Y.}~\bibnamefont {tuan
  Fang}}, \bibinfo {author} {\bibfnamefont {L.}~\bibnamefont {kun Chen}},
  \bibinfo {author} {\bibfnamefont {N.}~\bibnamefont {Zhu}}, \ and\ \bibinfo
  {author} {\bibfnamefont {J.}~\bibnamefont {Zhou}},\ }\href@noop {} {\bibfield
   {journal} {\bibinfo  {journal} {Optoelectronics, IET}\ }\textbf {\bibinfo
  {volume} {7}},\ \bibinfo {pages} {9} (\bibinfo {year} {2013})}\BibitemShut
  {NoStop}%
\bibitem [{\citenamefont {Sprung}\ and\ \citenamefont
  {Morozov}(2007)}]{Sprung:2007p3124}%
  \BibitemOpen
  \bibfield  {author} {\bibinfo {author} {\bibfnamefont {D.}~\bibnamefont
  {Sprung}}\ and\ \bibinfo {author} {\bibfnamefont {G.}~\bibnamefont
  {Morozov}},\ }\href@noop {} {\bibfield  {journal} {\bibinfo  {journal}
  {Journal of Physics A}\ } (\bibinfo {year} {2007})}\BibitemShut {NoStop}%
\bibitem [{\citenamefont {Hooper}\ \emph {et~al.}(2006)\citenamefont {Hooper},
  \citenamefont {Preist},\ and\ \citenamefont {Sambles}}]{Hooper2006}%
  \BibitemOpen
  \bibfield  {author} {\bibinfo {author} {\bibfnamefont {I.~R.}\ \bibnamefont
  {Hooper}}, \bibinfo {author} {\bibfnamefont {T.~W.}\ \bibnamefont {Preist}},
  \ and\ \bibinfo {author} {\bibfnamefont {J.~R.}\ \bibnamefont {Sambles}},\
  }\href@noop {} {\bibfield  {journal} {\bibinfo  {journal} {Phys. Rev. Lett.}\
  }\textbf {\bibinfo {volume} {97}},\ \bibinfo {pages} {053902} (\bibinfo
  {year} {2006})}\BibitemShut {NoStop}%
\bibitem [{\citenamefont {Butler}(2012)}]{butler}%
  \BibitemOpen
  \bibfield  {author} {\bibinfo {author} {\bibfnamefont {C.}~\bibnamefont
  {Butler}},\ }\emph {\bibinfo {title} {The Microwave Response of Square Mesh
  Metamaterials}},\ \href@noop {} {Ph.D. thesis},\ \bibinfo  {school}
  {University of Exeter} (\bibinfo {year} {2012})\BibitemShut {NoStop}%
\bibitem [{\citenamefont {McCalmont}\ \emph {et~al.}(1996)\citenamefont
  {McCalmont}, \citenamefont {Sigalas}, \citenamefont {Tuttle}, \citenamefont
  {Ho},\ and\ \citenamefont {Soukolis}}]{McCalmont}%
  \BibitemOpen
  \bibfield  {author} {\bibinfo {author} {\bibfnamefont {J.~S.}\ \bibnamefont
  {McCalmont}}, \bibinfo {author} {\bibfnamefont {M.~M.}\ \bibnamefont
  {Sigalas}}, \bibinfo {author} {\bibfnamefont {G.}~\bibnamefont {Tuttle}},
  \bibinfo {author} {\bibfnamefont {K.}~\bibnamefont {Ho}}, \ and\ \bibinfo
  {author} {\bibfnamefont {C.~M.}\ \bibnamefont {Soukolis}},\ }\href@noop {}
  {\bibfield  {journal} {\bibinfo  {journal} {Appl. Phys. Lett.}\ }\textbf
  {\bibinfo {volume} {68}} (\bibinfo {year} {1996})}\BibitemShut {NoStop}%
\bibitem [{\citenamefont {Griffiths}\ and\ \citenamefont
  {Steinke}(2001)}]{Griffiths:2001p2889}%
  \BibitemOpen
  \bibfield  {author} {\bibinfo {author} {\bibfnamefont {D.}~\bibnamefont
  {Griffiths}}\ and\ \bibinfo {author} {\bibfnamefont {C.}~\bibnamefont
  {Steinke}},\ }\href@noop {} {\bibfield  {journal} {\bibinfo  {journal} {Am.
  J. Phys.}\ }\textbf {\bibinfo {volume} {69}},\ \bibinfo {pages} {137}
  (\bibinfo {year} {2001})}\BibitemShut {NoStop}%
\bibitem [{\citenamefont {Butler}\ \emph {et~al.}(2009)\citenamefont {Butler},
  \citenamefont {Parsons}, \citenamefont {Sambles}, \citenamefont {Hibbins},\
  and\ \citenamefont {Hobson}}]{butler09}%
  \BibitemOpen
  \bibfield  {author} {\bibinfo {author} {\bibfnamefont {C.~A.~M.}\
  \bibnamefont {Butler}}, \bibinfo {author} {\bibfnamefont {J.}~\bibnamefont
  {Parsons}}, \bibinfo {author} {\bibfnamefont {J.~R.}\ \bibnamefont
  {Sambles}}, \bibinfo {author} {\bibfnamefont {A.~P.}\ \bibnamefont
  {Hibbins}}, \ and\ \bibinfo {author} {\bibfnamefont {P.~A.}\ \bibnamefont
  {Hobson}},\ }\href@noop {} {\bibfield  {journal} {\bibinfo  {journal} {Appl.
  Phys. Lett.}\ }\textbf {\bibinfo {volume} {95}},\ \bibinfo {eid} {174101}
  (\bibinfo {year} {2009})}\BibitemShut {NoStop}%
\bibitem [{\citenamefont {Butler}\ \emph {et~al.}(2011)\citenamefont {Butler},
  \citenamefont {Hooper}, \citenamefont {Hibbins}, \citenamefont {Sambles},\
  and\ \citenamefont {Hobson}}]{butler11}%
  \BibitemOpen
  \bibfield  {author} {\bibinfo {author} {\bibfnamefont {C.~A.~M.}\
  \bibnamefont {Butler}}, \bibinfo {author} {\bibfnamefont {I.~R.}\
  \bibnamefont {Hooper}}, \bibinfo {author} {\bibfnamefont {A.~P.}\
  \bibnamefont {Hibbins}}, \bibinfo {author} {\bibfnamefont {J.~R.}\
  \bibnamefont {Sambles}}, \ and\ \bibinfo {author} {\bibfnamefont {P.~A.}\
  \bibnamefont {Hobson}},\ }\href@noop {} {\bibfield  {journal} {\bibinfo
  {journal} {J. Appl. Phys.}\ }\textbf {\bibinfo {volume} {109}},\ \bibinfo
  {eid} {013104} (\bibinfo {year} {2011})}\BibitemShut {NoStop}%
\end{thebibliography}
\end{document}